\documentclass[conference]{IEEEtran}
\IEEEoverridecommandlockouts

\usepackage{cite}
\usepackage{amsmath,amssymb,amsfonts}
\usepackage{algorithmic}
\usepackage{graphicx}
\usepackage{textcomp}
\def\BibTeX{{\rm B\kern-.05em{\sc i\kern-.025em b}\kern-.08em
    T\kern-.1667em\lower.7ex\hbox{E}\kern-.125emX}}
\usepackage[nolist]{acronym}
\usepackage[draft]{pdfcomment}
\usepackage[dvipsnames]{xcolor}
\usepackage{orcidlink}

\begin{acronym}
\acro{THD}{total harmonic distortion}
\acro{CG}{current generator}
\acro{Bio-Z}{bio-impedance}
\acro{EIT}{electrical impedance tomography}
\acro{LUT}{lookup table}
\acro{DAC}{digital-to-analog converter}
\acro{HP}{half-period}
\acro{IC}{integrated circuit}
\acro{MASH}{multi-stage noise shaping}
\acro{PRNG}{pseudo-random number generator}
\acro{DWA}{data weighted averaging}
\acro{LSB}{least significant bit}
\acro{DEM}{dynamic element matching}
\acro{LPF}{low-pass filter}
\acro{SW}{switch}
\acro{RZ}{return-to-zero}
\acro{SFDR}{spurious-free dynamic range}

\end{acronym}
\hypersetup{hidelinks}      

\begin{document}
\title{\LARGE \textbf{A 0.5V, 6.2$\mu$W, 0.059mm\textsuperscript{2} Sinusoidal Current Generator IC\\with 0.088\% THD for Bio-Impedance Sensing}}
\author{
Kwantae Kim\orcidlink{0000-0001-8962-4554}, Changhyeon Kim, Sungpill Choi, and Hoi-Jun Yoo\orcidlink{0000-0002-6661-4879}\\
School of EE, KAIST, 291 Daehak-ro, Yuseong-gu, Daejeon 34141, Republic of Korea\\
Email: kwantae.kim@kaist.ac.kr
\thanks{This paper was accepted for publication and was presented in 2020 IEEE Symposium on VLSI Circuits (DOI: {\color{blue}\href{https://doi.org/10.1109/VLSICircuits18222.2020.9162983}{10.1109/VLSICircuits18222.2020.9162983}}). This preprint is intended to offer fully formatted list of citations, which was not feasible within the constraints of the original conference manuscript due to page limit.}
\thanks{\copyright 2020 IEEE. Personal use of this material is permitted. Permission from IEEE must be obtained for all other uses, in any current or future media, including reprinting/republishing this material for advertising or promotional purposes, creating new collective works, for resale or redistribution to servers or lists, or reuse of any copyrighted component of this work in other works.}}
\maketitle


\section*{\textbf{Abstract}}

This paper presents the first sub-10$\mu$W, sub-0.1\% \ac{THD} sinusoidal \ac{CG} \ac{IC} that is capable of 20kHz output for the \ac{Bio-Z} sensing applications. To benefit from the ultra-low-power nature of near-threshold operation, a 9b pseudo-sine \ac{LUT} is 3b $\Delta\Sigma$ modulated in the digital domain, thus linearity burden of the \ac{DAC} is avoided and only a 1.29$\mu$W of logic power is consumed, from a 0.5V supply and a 2.56MHz clock frequency. A \ac{HP} reset is introduced in the capacitive \ac{DAC}, leading to around 30dB reduction of in-band noise by avoiding the sampling of data-dependent glitches and attenuating the $kT/C$ noise and the non-idealities of reset switches (SW).


\section*{\textbf{Introduction}}

\ac{CG} of \ac{Bio-Z} sensor is an essential building block since the \ac{Bio-Z} is modulated into a voltage waveform by an injection current, so that it can be sensed by a voltage readout circuit. As a simple and low-power method of demodulation, chopping technique has been usually utilized in the voltage readout chain of the Bio-Z sensors, which in turn, eventually necessitates a low-distortion sinewave of the CG. For example, $<1$\% of \ac{THD} performance after chopping was used for the implantable pacemaker \cite{s.kim.isscc13}, while even much tighter \ac{THD} of $<0.5$\% at the \ac{CG} output was desired for the accurate \ac{EIT} applications \cite{m.kim.jssc17}. Yet, the generation of low-distortion ($<0.5$\%) sinewave has consumed a high-power ($>50\mu$W) in the prior arts \cite{m.kim.jssc17, s.-k.hong.tcas-ii19}, limiting the use of ultra-low-power applications (e.g. implantable) due to the severe lifespan reduction.

In this work, we propose a 6.2$\mu$W, 0.059mm\textsuperscript{2}, 0.088\% \ac{THD} sinusoidal CG with 20kHz and 2$\mu$App injection current. The key specification of low-distortion outperforms the recently published sub-10$\mu$W Bio-Z sensor \ac{IC} \cite{k.kim.isscc19} by 7.5$\times$, also exhibiting 8.9$\times$ less power consumption compared to the recently published sub-0.5\% \ac{THD} sinusoidal CG circuit \cite{s.-k.hong.tcas-ii19} at the same time. In addition, its compact chip area compared to the other sinusoidal \ac{CG} \acp{IC} allows the low-cost production.


\section*{\textbf{Proposed Sinusoidal \ac{CG} \ac{IC}}}

\begin{figure*}[t]
    \begin{center}
    \includegraphics[width=0.95\textwidth]{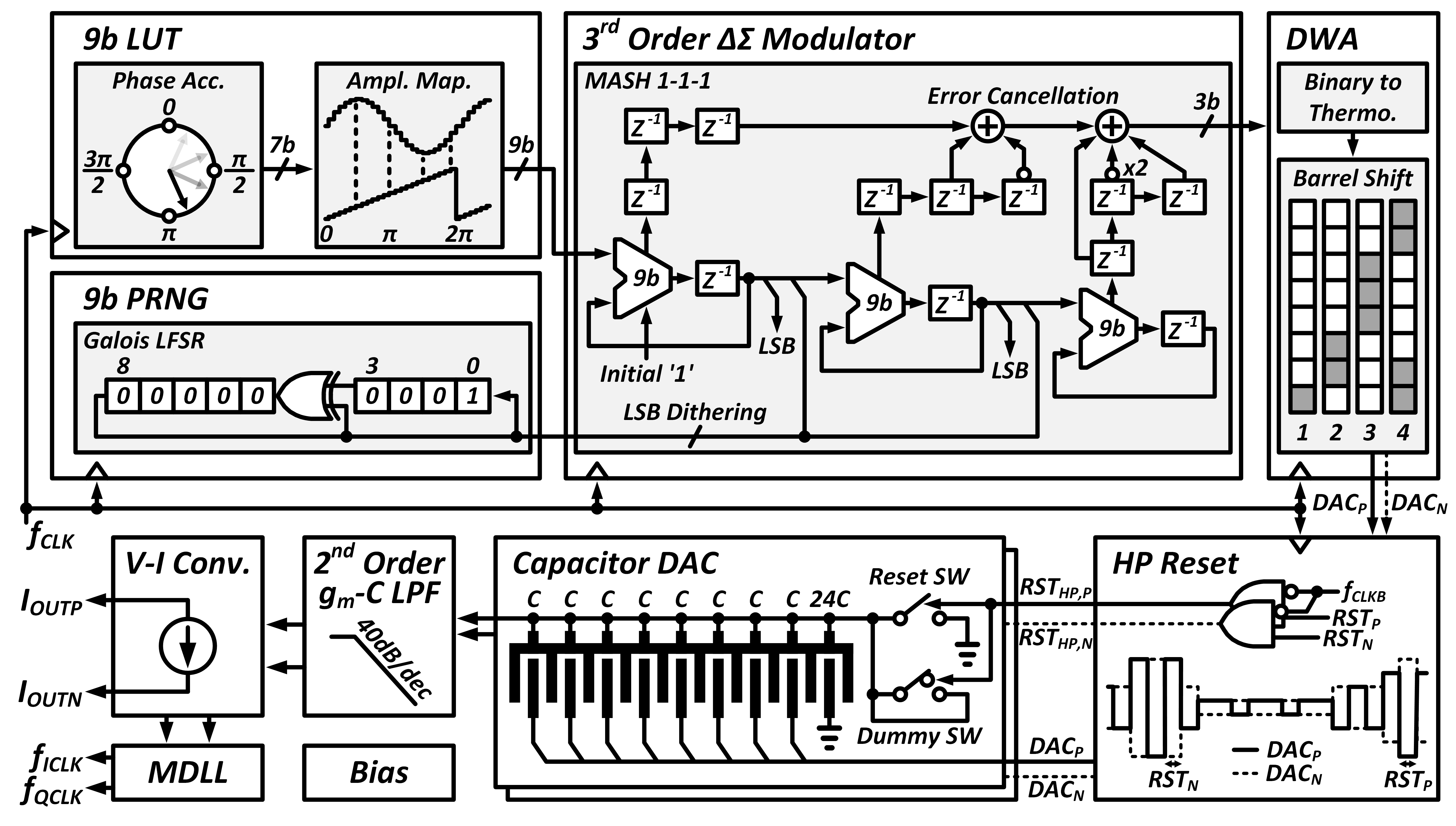}
    \caption{Overall architecture of the sinusoidal \ac{CG} \ac{IC}.}\label{fig:overall}
    \end{center}
    \vspace{-5mm}
\end{figure*}

The proposed low-distortion sinusoidal \ac{CG} \ac{IC} is shown in Fig.~\ref{fig:overall}. It includes digital/analog parts of building blocks all of which are fully operating in a low-supply voltage of 0.5V.

The digital part consists of a 9b pseudo-sine \ac{LUT}, a 3rd order digital \ac{MASH} 1-1-1 $\Delta\Sigma$ modulator, a 9b \ac{PRNG}, and a \ac{DWA}. When the phase accumulator sends a 128$\times$ rotational counter values to the amplitude mapped memory, the \ac{LUT} outputs a 9b integer-numbered sinewave data. The major challenge here to achieve sufficient spur reduction is the linearity burden of the \ac{DAC} because it is typically thermometer-coded in order to reduce the level-dependent glitches and matching requirements of the \ac{DAC} \cite{s.kim.isscc13, s.-k.hong.tcas-ii19}. Unfortunately, this approach results in a significant routing burden between the \ac{LUT} and the \ac{DAC} (512 wires), leading to a large area occupation not only due to the massive wires but also due to the pin allocation of \ac{LUT} during the logic synthesis. To cope with this implementation challenge, in this work, the digital $\Delta\Sigma$ modulation technique is adopted and it allows a significant area and design complexity reduction in both of the digital core and the DAC implementation, since only a 3b of thermometer-coded signals are interfacing with the \ac{DAC}. Although similar concept of approach was provided in \cite{s.-k.hong.tcas-ii19}, it required a large-bit-width of 18b in \ac{PRNG} to dither its poor spur performance of the oscillator, degrading its power efficiency. Here, the most of spur reduction is obtained only by the 9b \ac{LUT} that is sufficient to provide $>$70dB of in-band spur. Furthermore, the use of error-feedback-based $\Delta\Sigma$ modulator halves the required number of accumulators in each of the modulator stages compared to the output-feedback structure \cite{s.-k.hong.tcas-ii19}, greatly simplifying the implementation of the $\Delta\Sigma$ modulator. The \ac{PRNG} dithers the input \ac{LSB} of 2nd and 3rd accumulators which is also noise-shaped \cite{v.r.gonzalez-diaz.tcas-i10}, to further reduce the spurs. The \ac{DWA} block helps to enhance the linearity of \ac{DAC} by noise-shaped \ac{DEM} of the \ac{DAC} unit elements.

\begin{figure*}[t]
    \begin{center}
    \includegraphics[width=0.95\textwidth]{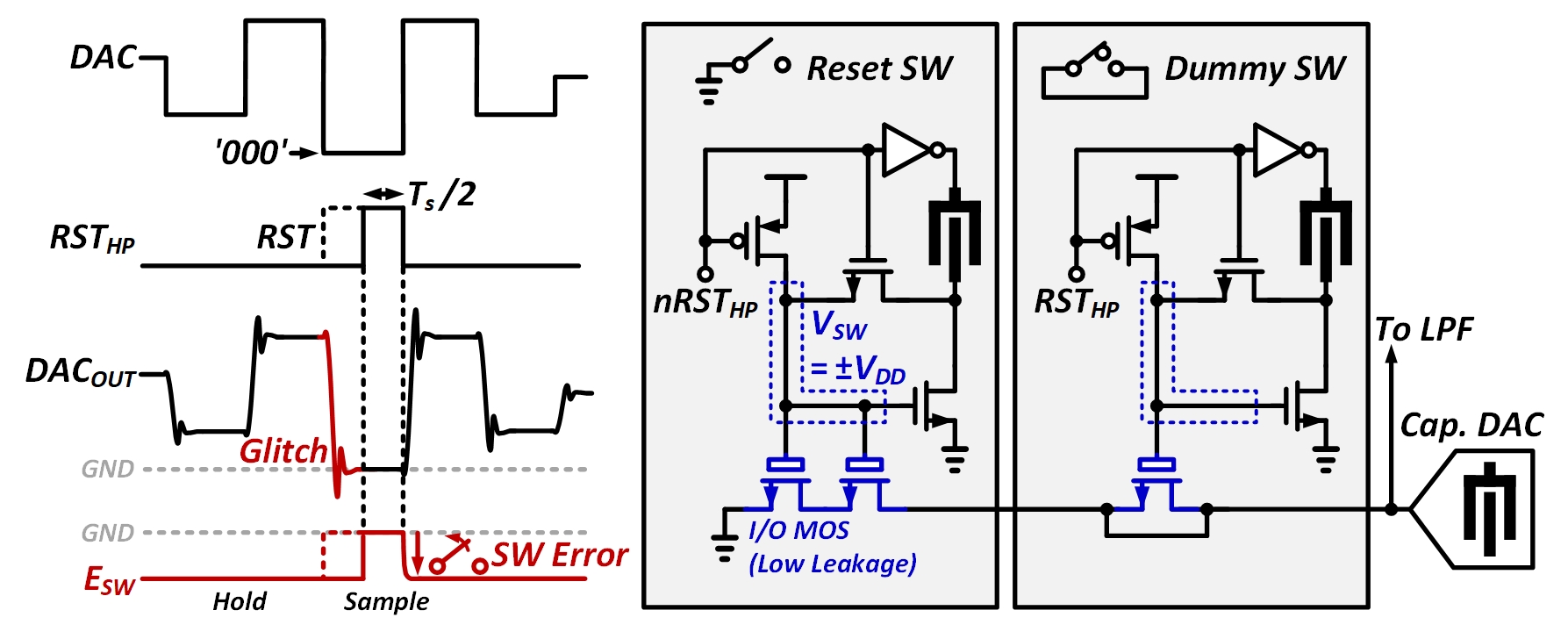}
    \caption{\ac{HP} reset.}\label{fig:hp_rst}
    \end{center}
    \vspace{-5mm}
\end{figure*}

The analog part consists of a 3b capacitive voltage \ac{DAC}, a 2nd order $g_{m}$-$\text{C}$ \ac{LPF}, and a V-I converter. The \ac{DAC} is implemented using the custom-layout metal-oxide-metal unit capacitors in a differential manner to achieve the low-power and even harmonic rejection. When the output of 3b $\Delta\Sigma$ modulated sinewave reaches ‘000’, the \ac{DAC} is reset to the GND. Since the reset instant in each of the $\Delta\Sigma$ modulated P/N \ac{DAC} output is not identical, not only the non-ideal errors of the reset \ac{SW} but also the data-dependent glitches which can be aliased when it is sampled, cannot be cancelled out by the differential structure. The consideration of glitches is especially important because the 3b $\Delta\Sigma$ modulated data sparsely reaches ‘000’, thus the frequency of reset operation is relatively low. In this work, we propose the use of \ac{HP} reset scheme in the capacitive \ac{DAC}. Rather than using the full period of clock for the duration of reset phase, a half-period is utilized using AND gates and the falling edge of the clock. The timing diagram of \ac{HP} reset is shown in Fig.~\ref{fig:hp_rst}. The \ac{HP} reset can avoid sampling of the glitches and the settling behavior due to the switching transients, and it halves the amplitude of not only clock feedthrough and charge injection errors but also the aliased amount of $kT/C$ noise in the frequency domain because the duty cycle of sample phase is halved, which is similar effect of the \ac{RZ} operation.

\begin{figure*}[t]
    \begin{center}
    \includegraphics[width=0.8\textwidth]{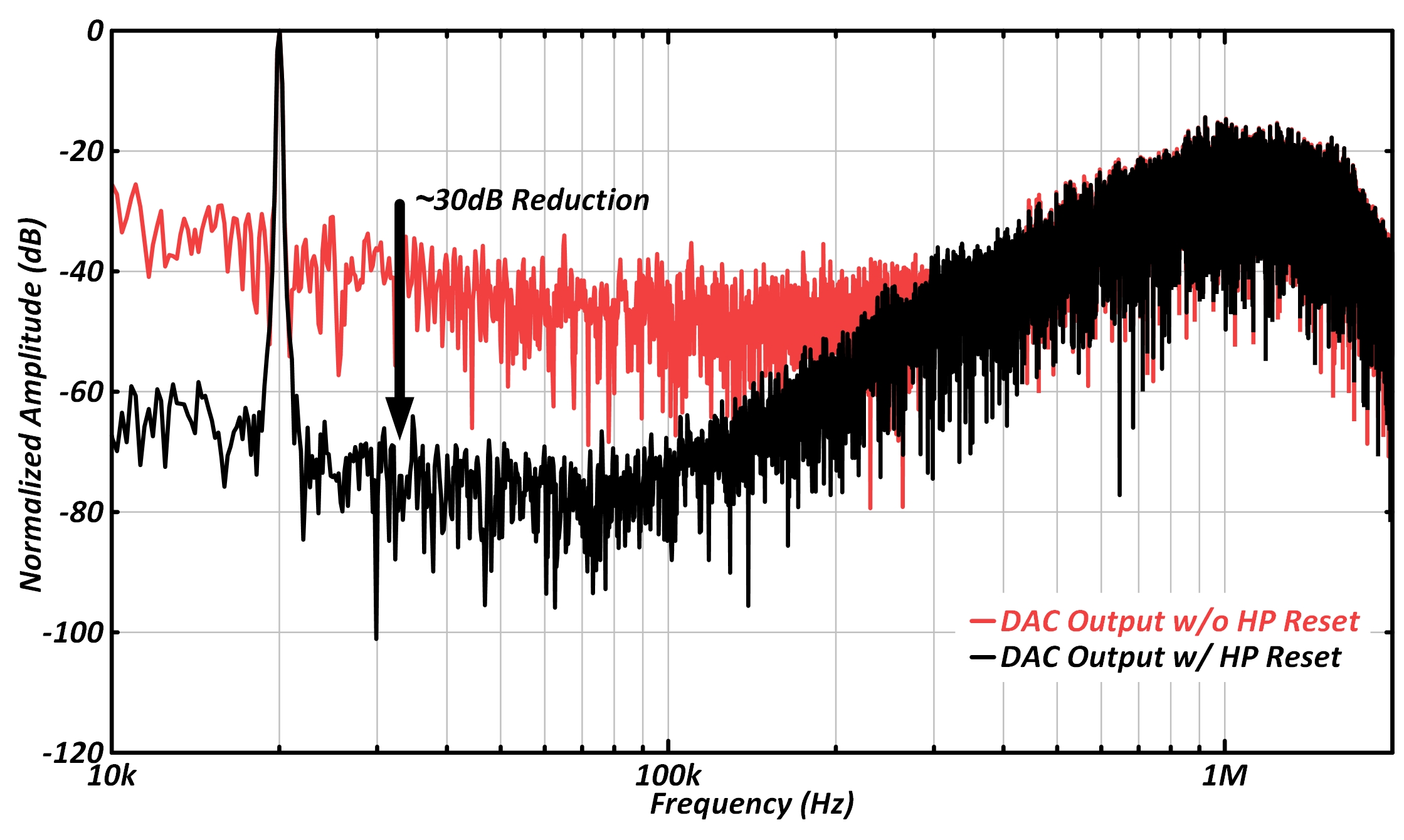}
    \caption{Simulated spectrum of the \ac{HP} reset scheme.}\label{fig:postlayout}
    \end{center}
    \vspace{-5mm}
\end{figure*}

As shown in Fig.~\ref{fig:postlayout}, the post-layout simulation result shows that a significant increase of in-band noise floor of the \ac{DAC} output can be avoided about 30dB with the help of \ac{HP} reset. Note that both of the simulation setup in Fig.~\ref{fig:postlayout} adopted dummy \ac{SW} to deal with the non-idealities by a first order. The implementation of reset \ac{SW} is shown in Fig.~\ref{fig:hp_rst}, based on the bootstrapped gate driving architecture of $\pm V_\text{DD}$ to suppress the leakage current.


\begin{figure*}[t]
    \begin{center}
    \includegraphics[width=0.93\textwidth]{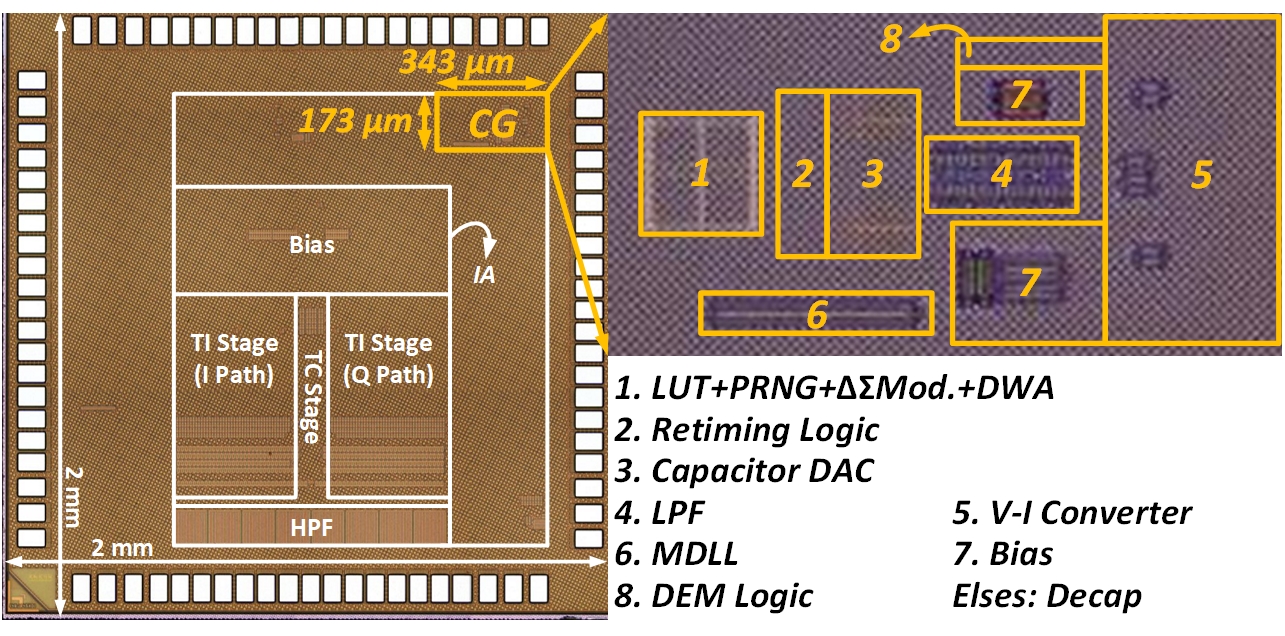}
    \caption{Chip photograph and the photograph of the same chip without top 3 layers (right).}\label{fig:chip_photo}
    \end{center}
    \vspace{-5mm}
\end{figure*}

\begin{figure*}[t]
    \begin{center}
    \includegraphics[width=0.8\textwidth]{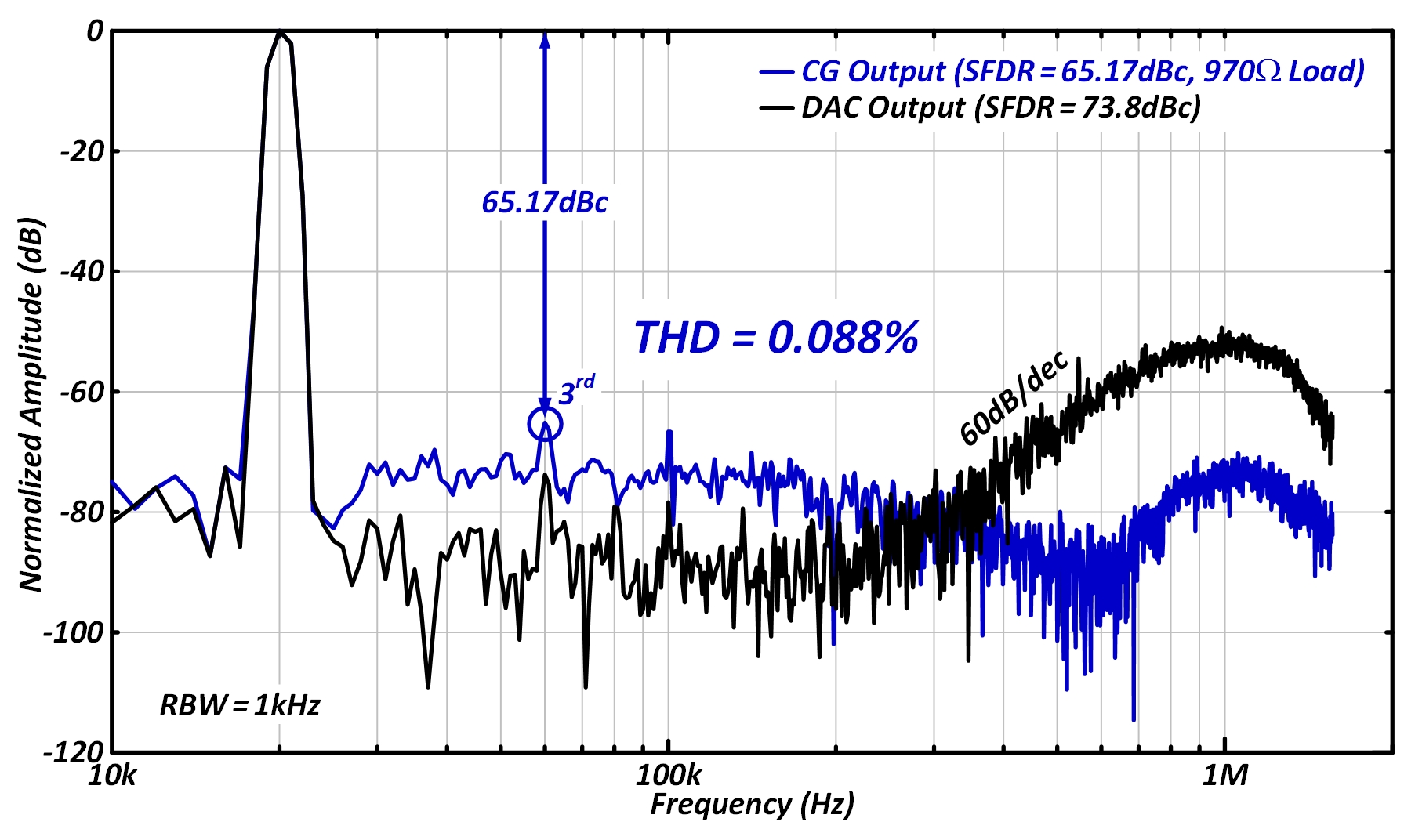}
    \caption{Measured spectrum of the \ac{CG} and the \ac{DAC} output.}\label{fig:mea_spectrum}
    \end{center}
    \vspace{-5mm}
\end{figure*}

\begin{figure*}[t]
    \begin{center}
    \includegraphics[width=0.8\textwidth]{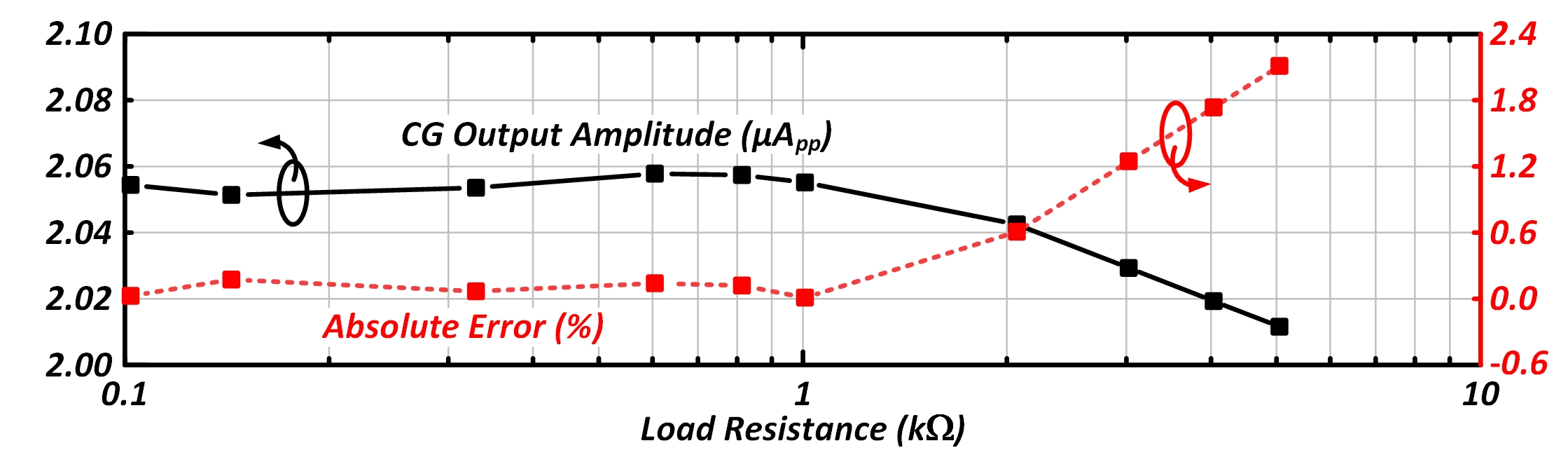}
    \caption{Measured load range the \ac{CG}.}\label{fig:mea_load}
    \end{center}
    \vspace{-5mm}
\end{figure*}

\begin{figure*}[t]
    \begin{center}
    \includegraphics[width=\textwidth]{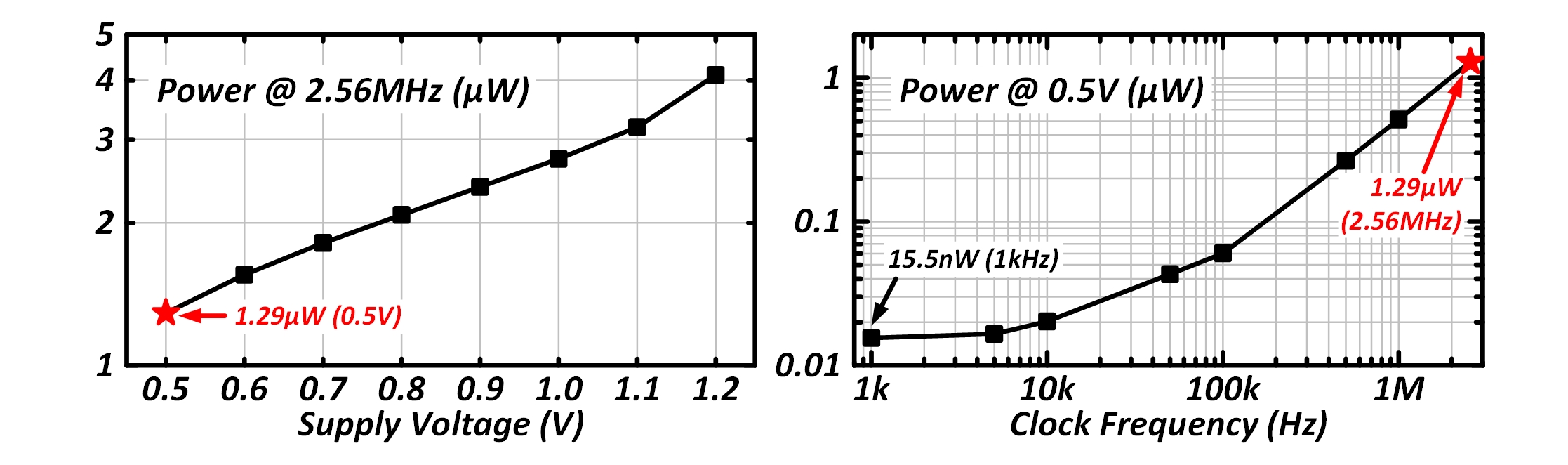}
    \caption{Measured logic power of the digital core.}\label{fig:mea_dig}
    \end{center}
    \vspace{-5mm}
\end{figure*}

\section*{\textbf{Measurement Results}}

\begin{figure*}[t]
    \begin{center}
    \includegraphics[width=\textwidth]{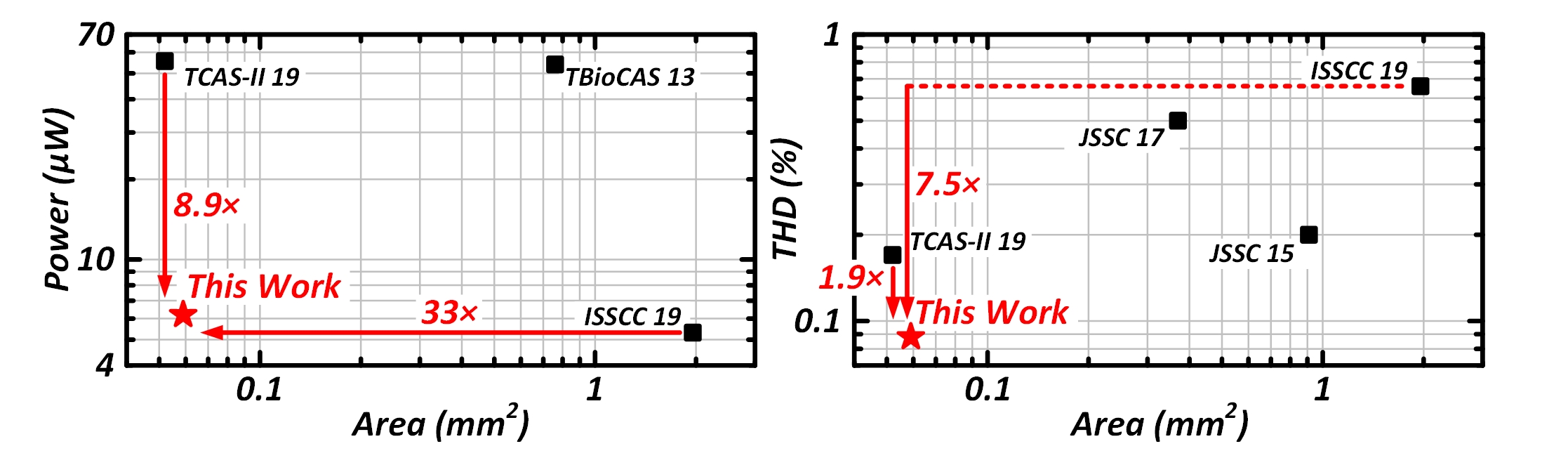}
    \caption{Performance comparison graph.}\label{fig:comparison}
    \end{center}
    \vspace{-5mm}
\end{figure*}

Fig.~\ref{fig:chip_photo} shows the proposed \ac{IC} which is fabricated in a 65nm CMOS, occupying only a 0.059mm\textsuperscript{2} of compact area. Fig.~\ref{fig:mea_spectrum} shows the measured spectrum of the 20kHz \ac{DAC} output and the 20kHz, 2$\mu$App sinusoidal current output of the proposed \ac{CG} \ac{IC}. Measured \ac{THD} and \ac{SFDR} performance of the \ac{CG} is 0.088\% (up to 20 harmonics) and 65.17dBc. The \ac{SFDR} is mainly bounded by the swing range of \ac{LPF} (10mV$_\text{pp}$ output) that is operating from the 0.5V of low-supply. Note that the residual high-pass-shaped 3rd order quantization noise is not completely removed due to the order of 2 of \ac{LPF}, however, it is out of bandwidth of the readout which is 400kHz for example in \cite{k.kim.isscc19}. Fig.~\ref{fig:mea_load} shows the measured load range of the \ac{CG} \ac{IC}, assuring $<$1\% of driving error up to 2k$\Omega$. Fig.~\ref{fig:mea_dig} shows the measured power consumption of the digital core. It consumes only a 1.29$\mu$W of operating power at 2.56MHz, with a leakage power of 15.5nW under the 0.5V supply. The digital core has a gate count of 1127. Table~\ref{fig:table} and Fig.~\ref{fig:comparison} summarizes the comparison of performances with the previous \ac{CG} \acp{IC} for the Bio-Z sensing applications. The proposed sinusoidal \ac{CG} \ac{IC} outperforms previous low-power design \cite{k.kim.isscc19} by 33$\times$ in chip area and by 7.5$\times$ in \ac{THD}. Furthermore, it also outperforms previous low-distortion design \cite{s.-k.hong.tcas-ii19} by 8.9$\times$ in total power and by 1.9$\times$ in \ac{THD}, recording the state-of-the-art low-power, low-distortion, and low-area sinusoidal \ac{CG} \ac{IC} for the \ac{Bio-Z} sensing applications.


\bibliographystyle{IEEEtran}
\bibliography{myref}

\begin{table*}[t]
    \begin{center}
    \caption{Performance summary and comparison.}\label{fig:table}
    \includegraphics[width=0.8\textwidth]{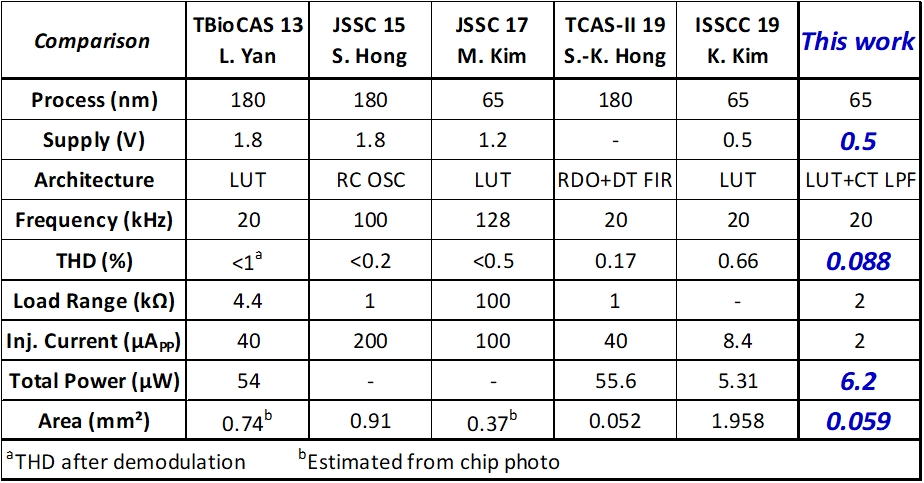}
    \end{center}
    \vspace{-5mm}
\end{table*}

\end{document}